\documentclass[aps,pre,epsfig,floats,twocolumn]{revtex4}   	

\usepackage{graphicx}				
\usepackage{amssymb}
\usepackage{amsmath}
\usepackage{hyperref}			
\usepackage{bm}				

\DeclareMathOperator{\sech}{sech}

\newcommand{\htyinf}[2]{$#1$-$#2$ healthy-infected}
\newcommand{\infhty}[2]{$#1$-$#2$ infected-healthy}
\newcommand{\toxofull}{{\textit{Toxoplasma gondii}}}
\newcommand{\toxo}{{\textit{T. gondii}}}

\newcommand{\bit}{\begin{itemize}}					
\newcommand{\eit}{\end{itemize}}					
\newcommand{\benum}{\begin{enumerate}}			
\newcommand{\eenum}{\end{enumerate}}				

\newcommand\be{\begin{equation}}							
\newcommand\ee{\end{equation}}							
\newcommand\bea{\be\begin{aligned}}						
\newcommand\eea{\end{aligned}\ee}  						

\newcommand{\figwidth}{\columnwidth}			

\newcommand{\eqRef}[1]{Eq. (\ref{#1})}				
\newcommand{\eqsRef}[1]{Eqs. (\ref{#1})}				
\newcommand{\eRef}[1]{(\ref{#1})}					
\newcommand{\figRef}[1]{Fig. \ref{#1}}				
\newcommand{\figsRef}[1]{Figs. \ref{#1}}				
\newcommand{\secRef}[1]{Sec. \ref{#1}}				

\newcommand{\citeRef}{Ref. }						
\newcommand{\citeRefs}{Refs. }					

\newcommand{\of}[1]{\left( #1 \right)}							
\newcommand{\set}[1]{\left\{ #1 \right\}}						
\newcommand{\avg}[1]{\left< #1 \right>}						

\begin{document}

\title{Hysteresis Effects in Social Behavior with Parasitic Infection}
\author{Michael Phillips}
\affiliation{Central New Mexico Community College, Albuquerque, New Mexico}
\date{14 November 2019 / Published: 9 June 2020}

\begin{abstract}
Recent work has found that the behavior of an individual can be altered when infected by a parasite.
Here we explore the question: under what conditions, in principle, can a general parasitic infection control system-wide social behaviors?
We analyze fixed points and hysteresis effects under the Master Equation, with transitions between two behaviors given two different subpopulations, healthy vs. parasitically-infected, within a population which is kept fixed overall. 
The key model choices are: (i) the internal opinion of infected humans may differ from that of the healthy population, (ii) the extent that interaction drives behavioral changes may also differ, and (iii) indirect interactions are most important. 
We find that the socioconfiguration can be controlled by the parasitically-infected population, under some conditions, even if the healthy population is the majority and of opposite opinion.
\end{abstract}

\maketitle


\section{Introduction}		\label{sec:intro}
Parasitic organisms, those which feed upon their hosts or use the hosts to obtain resources or to reproduce, often employ behavior-altering strategies such that the behavior of the infected host benefits the parasite more than the behavior of a healthy host\cite{parasite-animals,parasiterex,parasitebrain,toxohumans,socialflu,toxo-mice}. We use the term ``parasite" for any organism or virus taking residence in or on a host organism and evincing such a behavior-altering property during one or more stages of its life cycle.
Examples of these parasites are plentiful, ranging from viruses of the family \textit{Baculoviridae} (which affect caterpillars) to the fluke \textit{Dicrocoelium dendriticum} (which affects ants). 

In particular, research within the past two decades has uncovered significant influences on the behavior of various hosts from the protozoan parasite \toxofull. The infection, toxoplasmosis, has an acute phase followed by a latent (dormant) phase which can remain indefinitely. The latent phase of infection has no overt symptoms but significant behavioral differences have been observed, including examples in humans ranging from the occurrence of traffic accidents to entrepreneurship\cite{toxohumans,toxo-traffic1,toxo-traffic2,toxo-business,toxo-personality}. 

Extrapolating from the established behavioral changes of infected individuals, conjectures have been made about the ultimate implications for human activity and culture on a societal level\cite{parasiterex,parasitebrain,toxo-culture}. 
While it is clear that there should be significant macroscopic behavioral effects while the infected population holds the majority, the extent of macroscopic changes due to modest infected populations has not been assessed with mathematical rigor. Specifically, it is not clear if the role of social interactions will inhibit or exacerbate the aberrant behavior of infected individuals. We explore these macroscopic behavioral effects due to parasitic infection of some fraction of a large population of interacting individuals. 

We consider a very general and simple model (\secRef{sec:model}) which is not limited to any particular parasite or host, although one may keep in mind the prototypical example of \toxo\thinspace in humans. The population of host individuals is assumed to be very large, with each labelled according to their infection status and behavioral choice. 
Furthermore, the behavioral choice is assumed to be binary, which can be interpreted as a yes/no response or a two-party selection.
One detail which is not addressed here is that behavioral changes from parasitic infection can depend on gender\cite{toxo-gender}.

In \secRef{sec:results} we explore the behavioral dynamics with the infected population having an opposite behavioral preference to that of the healthy population, having a different inclination to agree with the majority behavior, and the combination of both of these together. In addition to the case of static parameters (\secRef{sec:staticpop}) we consider the effects of time-varying parameters, namely the generational variations of opinions (specifying behavioral preferences) of healthy individuals (\secRef{sec:oscopn}) and oscillations of the infected population as prevalence of the parasite in question varies over time
(\secRef{sec:oscpop}). 


\section{Model Details}	\label{sec:model}
We employ the Master Equation and approach the thermodynamic limit with a Fokker-Planck Equation, a method well-known from sociodynamics to produce reliable results\cite{weidlich1,helbing,haken-syn1,haken-syn2,castel-review,weidlich-pol,weidlich-migration,weidlich-qscience,weidlich-pscience,weidlich-stat,schweitzer-sim,ferm-approx}. 
We consider an overall population consisting of a fixed number of individuals, $N$. The population is then divided into two subpopulations: ``healthy" individuals (free of the parasite in question) denoted by $N_A$, and ``infected" individuals denoted by $N_B$: $N=N_A+N_B$. Each is split into two parts corresponding to the binary behavioral choices, labelled $j = 1, 2$:
\bea
\label{eq:popdefs}
N_{\mu} &= n_{\mu,1} + n_{\mu,2} \;\;\, \text{for} \;\; \mu = A,\, B		\\
N &= \sum_{\mu=A,B}\sum_{j=1,2} n_{\mu,j}	\, .	
\eea
The state of each individual is fully specified by the pair $\of{\mu,j}$, and the resulting socioconfiguration by the population in each state, $n_{\mu,j}$.

\subsection{Key Definitions}	\label{sec:defs}

The Master Equation governs the evolution of the probability distribution and involves transitions of individuals between the four states $\of{\mu,j}$. Transitions between subpopulations, $A \leftrightarrow B$, represent infection and cure processes which can be complex in nature and one-sided (dormant parasites may remain indefinitely)\cite{toxo}. This means the long-term configuration is largely infected unless birth/death processes are taken into account. To keep the model simple and investigate a range of infected populations, we reserve the ratio $x = \of{N_A-N_B}/N$ as a controlled parameter.	
Beginning with the four independent populations, $n_{\mu, j}$, we have introduced two constraints, $N$ and $x$, so two degrees of freedom remain. One of these is related to the difference of opinions $y$, while the other is a cross-difference $z$. With \eqsRef{eq:popdefs} we recast the four populations $n_{\mu, j}$ as the four variables,
\bea
\label{eq:vardefs}
	N &= n_{A,1} + n_{A,2} + n_{B,1} + n_{B,2}			\\
	x &= \frac{1}{N} \of{ n_{A,1} + n_{A,2} - n_{B,1} - n_{B,2} }		\\
	y &= \frac{1}{N} \of{ n_{A,1} - n_{A,2} + n_{B,1} - n_{B,2} }		\\
	z &= \frac{1}{N} \of{ n_{A,1} - n_{A,2} - n_{B,1} + n_{B,2} }		\,  ,
\eea
where only the last two enter as dynamic variables.
The variable $N \in \mathbb{R}^+$ has discrete spacing $\Delta N = 1$. The others $x,y,z \in \Gamma$, on the interval $[-1,1]$, have steps of $\delta x, \delta y, \delta z = 2/N \equiv \varepsilon$.

The configurational Master Equation then reads
\bea
 \label{eq:master1}
\frac{d}{dt} P(y,z;t) =  \sum_{y',z'\in\Gamma} & \left[		w^{z \leftarrow z'}_{y \leftarrow y'}(y',z') P(y',z';t) \right.	\\
& \left. - \,\; w^{z' \leftarrow z}_{y' \leftarrow y}(y,z) P(y,z;t)	\right]	.	
\eea
Considering only individual transitions, $n_{\mu,j} \to n_{\mu,j} \pm 1$, \eqsRef{eq:vardefs} imply $z'  = z \pm \delta z$ and $y' = y \pm \delta y$.
The four rates $w^{z' \leftarrow z}_{y' \leftarrow y}$ are related to the individual transition pobabilities $p_{\mu' j' \leftarrow \mu j}$,
 obtained by multiplying by the appropriate subpopulation, e.g. 
\be
\label{eq:trans}
w^{ z+\delta z \leftarrow z}_{ y+\delta y \leftarrow y } = n_{A,2} \, p_{A1 \leftarrow A2} = \frac{N}{4} (1+x-y-z) \, p_{A1 \leftarrow A2}	\,  .
\ee

Individuals in our model perceive (and are influenced by) the net behavior $y$ but cannot perceive (or don't care about) the relative population $x$ or the cross-difference $z$. 
Using an approach related to utility functions and focusing on indirect interactions, we denote the internal opinions (specifying behavioral preferences) of healthy and infected populations as $\alpha$ and $\beta$, respectively, the inclinations to behave like the majority as $\kappa_A$ and $\kappa_B$, and the rates of interaction as $\nu_A$ and $\nu_B$. The four individual transition probabilities per unit time are then written as:
\bea
\label{eq:indprobs}
p_{A1 \leftarrow A2}(y) = \nu_A \text{e}^{\alpha + \kappa_A y} \;\; ,& \;\; p_{A2 \leftarrow A1}(y) = \nu_A \text{e}^{-\alpha - \kappa_A y}	\\
p_{B1 \leftarrow B2}(y) = \nu_B \text{e}^{\beta + \kappa_B y} \;\; ,& \;\; p_{B2 \leftarrow B1}(y) = \nu_B \text{e}^{-\beta - \kappa_B y}	\; .	
\eea

\subsection{Mean-Field Drift Equations}	\label{sec:mean-field}

The Fokker-Planck formulation follows from the thermodynamic limit, $N \gg 1$ and each $n_{\mu,j} \gg 1$, where the step size $\varepsilon \ll 1$ allows for dynamical variables to be made continuous: $\delta y, \delta z \rightarrow dy, dz$ with $y, z \in \Gamma \to [-1,1]$. 
Specifically, it can be obtained via Taylor expansion for $(y', z')\approx(y, z)$ carried out to second order on the r.h.s. of \eqRef{eq:master1}. Drift is related to terms of first order in $\varepsilon$, leading to rescaled effective transition rates: $W^{ z\pm\delta z \leftarrow z}_{ y\pm\delta y \leftarrow y } \equiv \varepsilon \, w^{ z\pm\delta z \leftarrow z}_{ y\pm\delta y \leftarrow y }\;$\cite{weidlich1,gardiner}. Fluctuation dynamics are related to the second order terms $\sim \varepsilon^2 w^{ z\pm\delta z \leftarrow z}_{ y\pm\delta y \leftarrow y } = \varepsilon W^{ z\pm\delta z \leftarrow z}_{ y\pm\delta y \leftarrow y }$.

Thus, fluctuations are minimized by assuming a sharp unimodal initial distribution and taking a large $N \sim 1/\varepsilon$ \cite{ferm-approx}.
Alongside the fact that pair-wise and higher interactions are ignored, we can safely use the lowest order mean-value equations to assess the dynamics of this system rather than using the full distribution $P(y,z;t)$\cite{helbing,castel-review}. 
Evolutions of the mean values $\avg{y}_{t} = \int_{\Gamma^2} dy\,dz\,y\,P(y,z;t) \equiv y(t)$, and similarly $\avg{z}_{t} \equiv z(t)$, are then governed by the mean-field drift equations, obtained by integrating the Fokker-Planck expansion of \eqRef{eq:master1}:
\be
\label{eq:meandyn}
\frac{d}{dt} y(t) = K_y(y,z)	\; , \; \frac{d}{dt} z(t) = K_z(y,z) \, .
\ee

The drift coefficients are given by 
\bea
\label{eq:drifts}
K_y &= \sum_{m=\pm1} \of{W^{ z+m\delta z \leftarrow z}_{ y+\delta y \leftarrow y } - W^{ z+m\delta z \leftarrow z}_{ y-\delta y \leftarrow y } }	\\
K_z &= \sum_{m=\pm1} \of{W^{ z+\delta z \leftarrow z}_{ y+m\delta y \leftarrow y } - W^{ z-\delta z \leftarrow z}_{ y+m\delta y \leftarrow y } }	\, ,	
\eea
which can be written together as
\be
\label{eq:driftspm}
K_{y/z}(y,z) = \nu_A f_{+}(y,z;\alpha,\kappa_A) {\, \pm \,} \nu_B f_{-}(y,z;\beta,\kappa_B)
\ee
where
\be
\label{eq:driftfunc}
f_{\pm}(y,z;\eta,\kappa) = (1 \pm x) \sinh(\eta + \kappa y) - (y \pm z) \cosh(\eta + \kappa y)		\,	.
\ee


\section{Results}	\label{sec:results}

From \eqsRef{eq:indprobs} and the control of the relative population $x$, the system is controlled by 
seven independent parameters.
The two parameters, $\nu_A$ and $\nu_B$, govern the rates at which social interactions take place.
Since indirect interactions may be expected to occur with approximately equal frequency regardless of infection status we will set $\nu_A=\nu_B\equiv\nu$, and without further loss of generality we scale time so that $\nu = 1$.
The solutions of \eqsRef{eq:meandyn} are assessed in various regions of the remaining parameter-space: $( \alpha, \beta, \kappa_A, \kappa_B, x ) \in \mathbb{R}^4 \times \Gamma \equiv \mathcal{S}$. 	

We typically take $\beta < \alpha$ and $\kappa_B < \kappa_A$, i.e. the healthy population $A$ more strongly prefers behavior $1$, and more strongly seeks uniformity in behavior, than the infected population $B$. Taking distinct parameter values for healthy and infected subpopulations is the key element for the investigation of possible effects of parasites on the macroscopic socioconfiguration, $y$.

With all static parameters (\secRef{sec:staticpop}), at least one stable equilibrium $(y^{\ast}, z^{\ast})$ is guaranteed, corresponding to a peak in the stationary distribution $P_{0}(y,z)$ \cite{weidlich1,helbing,arfken,gardiner}. When one personal parameter is allowed to vary (\secRef{sec:oscopn}), the effects of a static relative population are combined with those of a varying personal trait.

If the relative population $x$ varies on shorter time-scales than oscillations of personal opinions, then the latter can be considered static parameters (\secRef{sec:oscpop}). This possibility allows for non-standard hysteresis effects: the socioconfiguration at one time may be decided by the relative population that existed at some earlier time. 

We consider several cases with dynamically varying parameters driving macroscopic behavioral changes, in an attempt to probe the large parameter space.

\subsection{Static Parameters}	\label{sec:staticpop}

The simplest case of static parameters refers to a single point in $\mathcal{S}$. 
According to the drifts, \eqsRef{eq:meandyn} with \eqsRef{eq:driftspm} and \eRef{eq:driftfunc}, we find the fixed points
\bea
\label{eq:general-fp}
y^{\ast} &= \frac{1}{2} \left[ \of{1+x} \tanh(\alpha + \kappa_A y^{\ast}) + \of{1-x} \tanh(\beta + \kappa_B y^{\ast})	 \right]	\\
z^{\ast} &= \of{1+x} \tanh(\alpha + \kappa_A y^{\ast}) - y^{\ast}	.	
\eea
The cross-difference reaches a fixed point $z^{\ast}$ completely determined by that of the behavior $y^{\ast}$. This justifies focusing on the behavior $y$ in the following, and the cross-difference $z$ can be inferred from \eqsRef{eq:general-fp} if desired (approximately, when a parameter is oscillating).

The transcendental form of \eqsRef{eq:general-fp} makes the fixed points analytically inaccessible. 
However, it is clear that the number of fixed points is 1, 2, or 3; up to two will be stable. 
A transition occurs as the r.h.s. of \eqRef{eq:general-fp} intersects the l.h.s. tangentially, where a new fixed point emerges or switches stability. If the emerging fixed point is small, $y^{\ast} \approx 0$, we find a transition condition
\be
\label{eq:crossover}
(1+x) \sech^{2}(\alpha) \kappa_A + (1-x) \sech^{2}(\beta) \kappa_B = 2		\,	.
\ee
The original fixed point, $(y^{\ast},z^{\ast}) = (0,0) \text{ if }\beta = - \alpha \text{ and } x=0$, becomes unstable when the l.h.s. of \eqRef{eq:crossover} surpasses $2$, giving rise to a pair of stable fixed points away from $(0,0)$ despite the counter-acting individual opinions. This is reminiscent of typical symmetry-breaking transitions, e.g. a net magnetization arises (in a direction governed by the initial condition) when the coupling parameter crosses a critical value\cite{kardar-f}.

\begin{figure}[h]
\includegraphics[width=\figwidth]{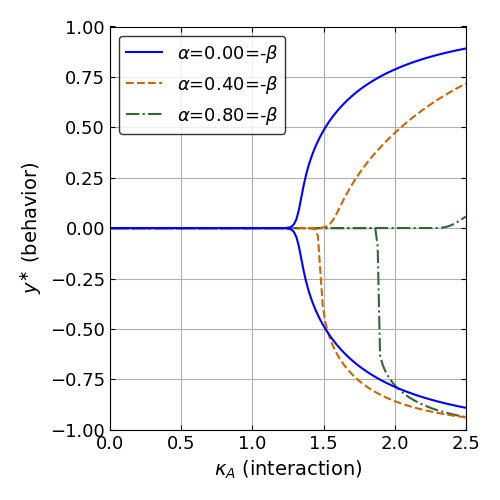}
\caption{Phase diagram: the fixed point $y^{\ast}$ shown as a function of the interaction parameter $\kappa_A$. The other interaction is not fixed but follows $\kappa_B = \kappa_A / 2$, while the remaining parameters are: $x = 0, \, \alpha = -\beta \in \set{0,0.4,0.8}$. Asymmetric bifurcations arise from unequal interactions with polar opinions: fixed points favoring behavior 2 ($y^{\ast}<0$) appear as healthy individuals adapt to the status quo due to $\kappa_A$, while those favoring behavior 1 ($y^{\ast}\gtrsim0$) rely on infected individuals falling in line due to $\kappa_B$.}
\label{fig:phase-trans}
\end{figure}

\begin{figure}[h]
\begin{center}
\begin{tabular}{l | l}
\includegraphics[width = 0.575 \columnwidth]{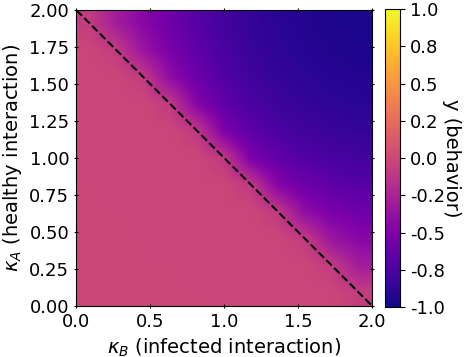}		&
\includegraphics[width = 0.57 \columnwidth]{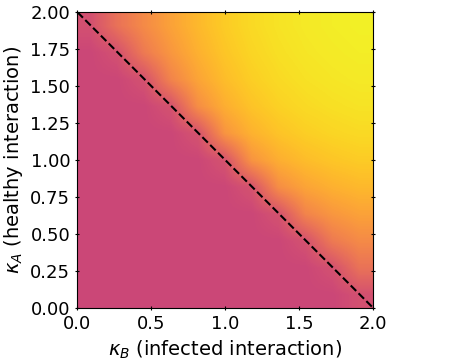}	
\end{tabular}
\end{center}
\caption{Behavior $y^{\ast}$ as a function of the interaction parameters $\kappa_A, \kappa_B$, for initial conditions $y(0) < 0$ (left) and $y(0) > 0$ (right).
Here: $x=0, \, \alpha=\beta=0$.
The dashed line shows $\kappa_A + \kappa_B = 2$.		
With sufficiently strong interactions $\kappa_A + \kappa_B > 2$, behavior 2 dominates for $y(0) < 0$ while behavior 1 dominates for $y(0) > 0$.	\newline
[An interpolation is applied between fixed points $y^{\ast}(\kappa_A, \kappa_B)$ on a discrete $10\times10$ grid.]
}
\label{fig:2dphase0}
\end{figure}

Like many symmetry-breaking transitions it is marked by a pitchfork bifurcation, which can be found by numerically solving \eqsRef{eq:meandyn} and extracting the fixed points for different initial conditions\cite{strogatz}.
When the populations are balanced, $x = 0$, with neutral opinions, \eqRef{eq:crossover} predicts a transition near $\kappa_A + \kappa_B = 2$. If infected individuals only consider interactions favoring behavior 2, with the same weight as healthy individuals consider either interaction, then $\kappa_B = \kappa_A / 2$ and the transition occurs near $\kappa_A = 4/3$. 
This is confirmed by numerical calculation, as seen in \figRef{fig:phase-trans}. The boundary near $\kappa_A + \kappa_B = 2$ is revealed more generally by allowing both parameters to vary independently, shown in \figRef{fig:2dphase0}.

Bifurcations for opposite opinions $\alpha = -\beta \neq 0$ are numerically found to occur at different values of $\kappa_A$, the shift toward behavior 1 ($y^{\ast}>0$) relying upon transitions of the infected subpopulation via $\kappa_B$. The critical value predicted by \eqRef{eq:crossover} corresponds to the emerging fixed point $y^{\ast} \gtrsim 0$, leading to $\kappa_A \approx 1.56$ for $\alpha = -\beta = 0.4$ and $\kappa_A \approx 2.38$ for $\alpha = -\beta = 0.8$ (with $\kappa_B=\kappa_A/2$), and is preceded by the emergence of the fixed point $y^{\ast} < 0$ as seen in \figRef{fig:phase-trans}. 
The condition for these latter fixed points is not analytically available because they originate from tangential intersections of the l.h.s. and r.h.s. of \eqRef{eq:general-fp} away from $y^{\ast}=0$.  
Phase diagrams like \figRef{fig:2dphase0} for opposite opinions merely have a shifted offset of the transition line with the slope unaffected, while those for imbalanced populations $x\neq0$ can be more complex.

\subsection{Oscillating Opinions}	\label{sec:oscopn}

We assess the effects of the infected population while the opinion of the healthy population (the ``healthy opinion") is varying in time. Specifically, we allow for oscillations due to generational differences in opinion, so that line segments in $\mathcal{S}$ are explored (this aspect is similar to a model from \citeRef\cite{weidlich1}).
All parameters may generally vary with time --- here we assume the time-scale of opinion changes is much shorter than that of population changes, with the shortest time-scale determined by the interaction rate: $1/\nu < 2\pi/\omega_\alpha \ll 2\pi/\omega_x$.

We consider a neutral infected opinion, $\beta=0$, and an infected opinion favoring behavior 2, $\beta=-0.4$. Each is applied to relative populations $x\in\{0,\pm0.4\}$, $50$-$50$ and $70$-$30$ splits of the population. This approaches the maximum $80$-$20$ split for latent toxoplasmosis\cite{toxo}.

\begin{figure}[h]
\includegraphics[width=\figwidth]{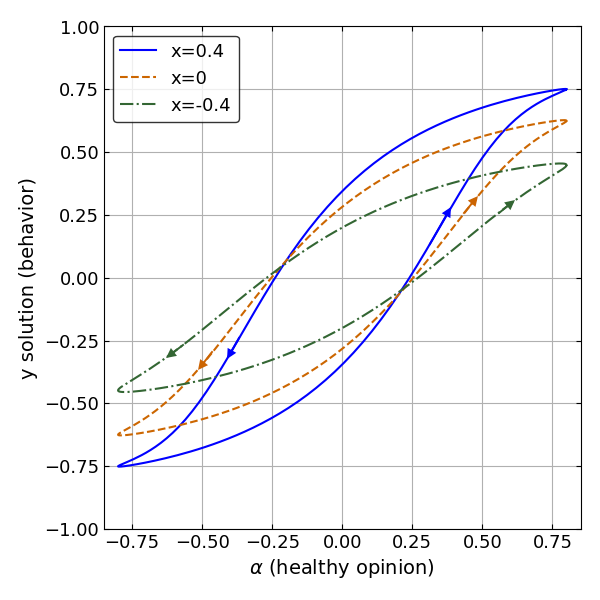}
\caption{Solution $y(t)$ plotted against the varying opinion $\alpha(t)\in[-0.8,0.8]$ for $\beta=0$ and $\kappa_A=\kappa_B=0.7$. The initial path is unimportant and is omitted, leaving just a closed hysteresis curve for each relative population $x$. }
\label{fig:sol-xp}
\end{figure}

\figRef{fig:sol-xp} shows hysteresis loops for $\beta=0$, from numerical solutions of \eqsRef{eq:meandyn} with an oscillating opinion $\alpha(t) = \alpha_0 \sin(\omega_{\alpha}t)$. 
We take $\omega_{\alpha}= 0.25$, corresponding to a period $\approx25/\nu$, and $\alpha_0 = 0.8$, a strong maximum intra-generational agreement in opinion.
Interaction parameters have little qualitative effect within the interval $\kappa_{A,B} \in [0,1.5]$, so we use moderate values $\kappa_A = \kappa_B = 0.7$. Larger $\kappa_{A,B}$ as in \eqRef{eq:crossover} gives wider and steeper loops, reflecting features of a hard ferromagnet, which still span negative and positive values.

\begin{figure}[h]
\includegraphics[width=\figwidth]{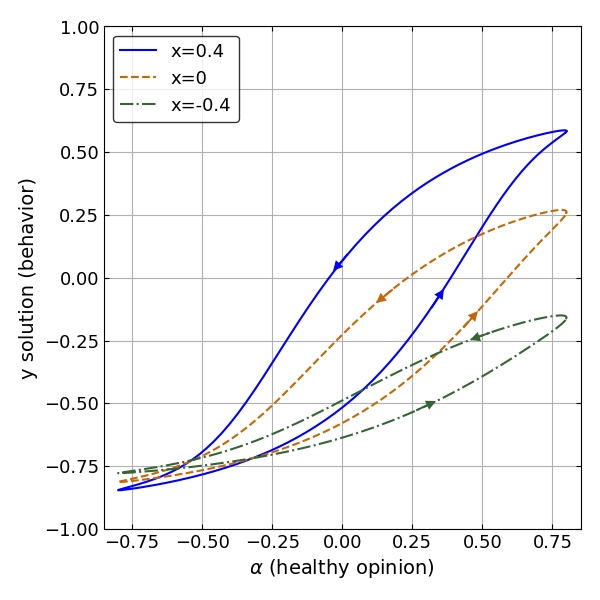}
\caption{Hysteresis curves as in \figRef{fig:sol-xp} for $\beta=-0.4$ and $\kappa_A = \kappa_B = 0.7$.
When the majority is infected ($x<0$), the net behavior always agrees with their opinion ($y<0$ always).}
\label{fig:sol-x0xm}
\end{figure}

Hysteresis loops for $\beta=-0.4$ are shown in \figRef{fig:sol-x0xm}.
A $50$-$50$ split of the population allows for behaviors $y>0$ only for $\alpha\gtrsim0.2$, while a \infhty{70}{30} split keeps $y<0$ for all $\alpha\in[-0.8,0.8]$. 
This suggests a phase transition occurs when the infected population crosses a critical value, along with a moderate infected opinion, in contrast to the typical transition caused by an increase of the interaction parameter.

\subsection{Oscillating Populations}	\label{sec:oscpop}

Here we consider the case of an oscillating relative population, exploring line segments in $\mathcal{S}$ orthogonal to those of \secRef{sec:oscopn}.
We now assume the time-scale of changes in relative population is much shorter than that of changes in other parameters, while the shortest time-scale is still given by the interaction rate: $1/\nu < 2\pi/\omega_x \ll 2\pi/\omega_\alpha$. Specifically, we take $x(t) = x_0 \sin(\omega_{x} t)$ with $\omega_{x} = 0.25$ and $x_0 = 0.6$, allowing for an evolution with extrema representing $80$-$20$ splits of the population. 

We consider the following cases: a neutral healthy opinion with a moderate infected opinion, $\alpha = 0$ and $\beta = -0.4$, for a selection of interactions on the line $\kappa_A + \kappa_B = 1.4$; and opposite opinions, $\alpha = 0.4$ and $\beta = -0.4$, with interactions $\kappa_A \geq \kappa_B$ on the line $\kappa_A + \kappa_B = 2.2$. 

These cases illustrate how the net behavior can be controlled by the infected population and/or opinion to varying degrees. The neutral healthy opinion necessarily leads to net behaviors in favor of the infected opinion, while the opposite opinions typically keep the net behavior centered near zero. The interaction parameters are chosen such that regions before and after the phase transition are explored (see \figsRef{fig:phase-trans}, \ref{fig:2dphase0}).

\begin{figure}[h]
\includegraphics[width=\figwidth]{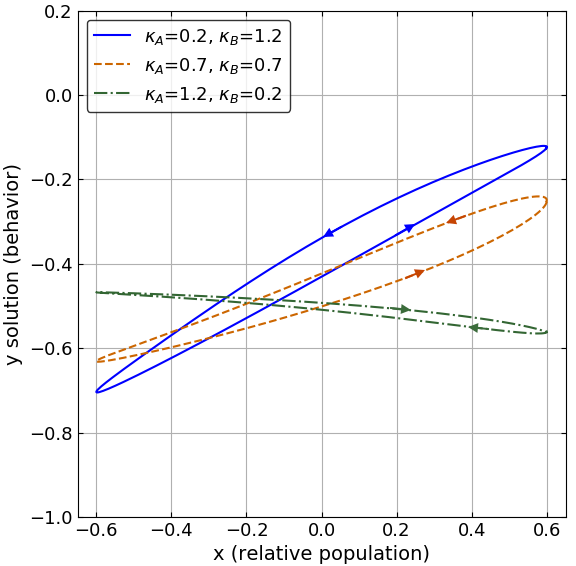}
\caption{Net behavior $y(t)$ with parameters $\alpha=0, \beta=-0.4$, and $\kappa_A + \kappa_B = 1.4$. Soft hysteresis loops result, but the loop for $\kappa_A=1.2$ is unlike the others: its axis is sloped downward, and it is traversed in the opposite direction (clockwise). In that case, the infected behavior is adopted more widely as the infected population \emph{decreases}, due to the pressure among healthy individuals to adopt the established majority behavior and increase uniformity.}
\label{fig:sol-kmed}
\end{figure}

\figRef{fig:sol-kmed} reveals all hysteresis loops with $\alpha=0$ centered near $y \approx -0.5$. 
The loops reflect a soft hysteresis, due to the moderate choice $\kappa_A + \kappa_B = 1.4$. 

Interestingly, a large $\kappa_A=1.2$ (small $\kappa_B=0.2$) leads to a loop with its long axis making a slope opposite to the others, and traversed in the opposite direction. The net behavior agrees more strongly with the preference of the infected population when that population is a \emph{minority} --- even though healthy individuals are neutral in opinion, they are much more inclined to agree with the status quo than the infected individuals who, in contrast, act almost independently. So the net behavior takes more negative values as $x$ increases, as more (healthy) individuals fall in line with the established (infected) behavior.

\begin{figure}[h]
\includegraphics[width=\figwidth]{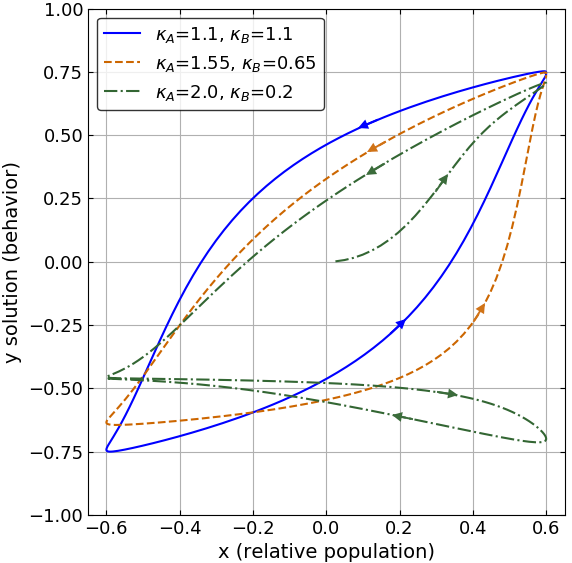}
\caption{Behavior $y(t)$ with $\alpha=0.4, \beta=-0.4$, and $\kappa_A + \kappa_B = 2.2$. Hard hysteresis loops result for $\kappa_A \leq 1.55$. The initial evolution is retained for $\kappa_A=2$ since the system is later trapped in a loop with $y<0$. This case resembles the down-sloped loop of \figRef{fig:sol-kmed}, again due to the large social pressure for uniformity among healthy individuals. }
\label{fig:sol-klrg}
\end{figure}

For the case of opinionated healthy individuals, $\alpha=0.4$, with a counteracting infected opinion, the net behavior typically agrees with the majority population from a previous time due to the delay from interactions, shown in \figRef{fig:sol-klrg}. A symmetric and comparatively hard hysteresis loop arises for $\kappa_A = \kappa_B = 1.1$, whereas an asymmetric loop, made steeper as the healthy population increases and less steep as the infected population increases, arises for distinct interaction values $\kappa_A > \kappa_B$. 

A completely different evolution is uncovered for $\kappa_A \gg \kappa_B$. The initial evolution from $y(0)=0$ is retained, and mimics the saturation and soft reversal of other loops, but the behavior is thereafter confined to a loop about $y \approx -0.6$. The loop is traversed in the opposite direction and its axis is sloped downward, which bears similarity to the case of \figRef{fig:sol-kmed} with $\kappa_A>\kappa_B$ except the net behavior now agrees more strongly with the infected opinion despite the opposing healthy opinion. The very large value $\kappa_A = 2$ has many healthy individuals insisting upon agreement with the pre-existing majority behavior, even when the vast majority is healthy (\htyinf{80}{20}) and quite opposed to that behavior. This point is similar to the minority-rule case from \citeRef \cite{gob-opndyn} and the ``spiral of silence" from \citeRefs \cite{nn-spiral1,nn-spiral2}.

If the opinions are made significantly more polar, $\alpha = -\beta > 0.4$, the loop for $\kappa_A=2$ becomes similar to the others.
Such strong global opinions may not be reasonable due to biological and social fluctuations.	


\section{Discussion}	\label{sec:disc}

The net behavior can favor one side over the other due to an asymmetry in the parameters, or due to a symmetry-breaking transition under symmetric parameters. Either way, any behavior away from $y=0$ can emerge even when the difference in opinions is modest, $|\alpha -\beta| \lesssim 1$. The system is trapped with $y<0$ when the infected population is the majority, even with moderate interactions $\kappa_A + \kappa_B \sim 1.4$. This also occurs when the infected population is smaller or dynamic, but only if interactions are sufficiently large and far apart: $\kappa_A + \kappa_B \gtrsim 2$ and  $\kappa_A \ll \kappa_B$ or $\kappa_A \gg \kappa_B$.

The conjectured far-reaching societal effects arising from toxoplasmosis\cite{parasiterex,parasitebrain,toxo-culture} are possible, although we found that the infected population must be the majority at some moment for it to be impactful. The role of interactions has given rise to delays from hysteresis: the net behavior now reflects the population and opinion split that was present some time ago. This delay grants some power to the infected opinion since the net behavior can be controlled by the appearance of a majority infected population, even if it doesn't persist at all.

The most extreme case of the infected opinion taking control of the system is when one interaction parameter is so large that it amounts to authoritarian pressure for uniformity. Once the net behavior favors the infected opinion even slightly, the system is forever trapped in favor of it indefinitely, and actually worsens when the infected population decreases. This extreme case is perhaps unrealistic, unless a strict authoritarian uniformity is enforced by government or culture, but it shows the point at which interactions grossly exacerbate the presence of an infected population. 

The simple model employed in this work cannot be trusted for all details, but it is sufficiently general that it can be applied to any problem with two different subpopulations of individuals undergoing transitions between binary behaviors\cite{flache-timing,fu-hkmodel,gob-opndyn,bnr-diversity,mw-factors}. It could be applied to explore differences in opinions and interactions due to gender, race, economic background, or any dyad of populations. The form of Eqs. (\ref {eq:indprobs}) is also equivalent to a physical problem of mean-field magnetism, with two species of particles (spins) which respond to distinct magnetic fields $\sim \alpha , \beta $, and respond differently to the mean field $y$.



\end{document}